# Losses in Superconductors under Non-Sinusoidal Currents and Magnetic Fields

Vladimir Sokolovsky, Victor Meerovich, Marat Spektor, George Levin, and Istvan Vajda

***Abstract*—Study of AC losses in superconducting wires and tapes is usually restricted by consideration of applied sinusoidal currents and/or magnetic fields. However, currents in electric power systems contain a wide variety of harmonics. The currents become strongly non-sinusoidal at the operation of converters, non-linear reactors, and during transient and overload conditions. We report the results of the analysis of the influence of higher harmonics of the current and magnetic field on AC losses in superconducting slabs, strips and coated conductors. Analytical expressions are obtained in the framework of Bean's critical state model; the power law voltage-current characteristics are treated numerically. It is shown that the contribution of higher harmonics to AC losses in superconducting elements can be tens times larger than in normal metals and the 5% harmonic can increases the losses by up to 20%.**

## I. INTRODUCTION

AC loss values determine the range of the rated currents and magnetic fields for superconducting devices, required power of cryogenic equipment, and economical gain. Numerous investigations have been devoted to consideration of AC losses in superconductors of various form, wires, tapes, coated conductors, etc. under different conditions (see for example [1-6] and references therein). However, most of the theoretical works is restricted to sinusoidal magnetic fields or/and currents. In reality, currents $I(t)$ in electric power systems contain a set of harmonics and can be presented as

$$I(t) = \sum_k I_k \cos(k\omega t + \varphi_k) \tag{1}$$

where $\omega = 2\pi f$, $f$ is the frequency of the main harmonic, $I_k$ and $\varphi_k$ are the amplitude and phase of the $k$-th harmonic, ($k$ = 0,1,2,…). The frequency $f$ for power systems is 50 or 60 Hz, for special electric systems (airplanes, ships, etc.) frequently is 400-800 Hz.

For normal metal parts of a device the Joule losses $P_j$ caused by a current $I(t)$ and the eddy losses $P_e$ caused by a magnetic field $H(t)$ are presented in the form:

$$P_j = \frac{R(f) I_1^2}{2}\left(1 + \sum_{k \neq 1} r_k i_k^2\right) \tag{2a}$$

$$P_e = \frac{K_1(f) H_1^2}{2}\left(1 + \sum_{k \neq 1} p_k h_k^2\right) \tag{2b}$$

where $R$ is the conductor resistance at the frequency $f$, $r_k$ give the frequency dependence of the resistance, $i_k = I_k/I_1$, $h_k = H_k/H_1$, $H_k$ is the $k$-th harmonic amplitude of the magnetic field, $K_1(f)H_1^2/2$ is the eddy losses caused by the main harmonic, $p_k$ characterize the dependence of these losses on frequency. The explicit forms of $R$, $K_1$, $r_k$, $p_k$ depend on the conductor shape and magnetic field direction. For example, for losses per unit of length in a thin normal metal strip in magnetic field directed perpendicular to wide surface

$$K_1 = \frac{2\mu_0^2 w^2 a \omega^2}{3\rho} \text{ and } p_k = k^2 \tag{3}$$

where $\mu_0$ is the magnetic permeability of the vacuum, $\rho$ is the resistivity, $a$ is the strip thickness, $w$ is the a half of the strip width.

According the requirements to power quality supported by power systems, current waveforms have to be very close to sinusoidal. The amplitudes of the higher harmonics should not exceed a few percents of the main harmonic amplitude. Usually the contribution of higher harmonics decreases rapidly, at least as $1/k$, and it is sufficiently to take into account only several first harmonics. $I_0$ = 0 because a direct current cannot be transferred by a transformer. So, with the accuracy of about 1%, the losses in normal conducting parts of power devices are determined by the main harmonic. Currents become strongly non-sinusoidal at the operation of converters, non-linear reactors, and during transient and overload regimes. In this case all harmonics have to be taken into account at the loss estimation. Since superconductors possess a strongly non-linear current voltage characteristic, one can expect substantial contribution of higher harmonics to the AC losses in superconducting elements.

In this paper we analyze the influence of higher harmonics on AC losses in superconductors.

V. Sokolovsky, V. Meerovich, and M. Spektor are with the Physics Department, Ben Gurion University of the Negev, P.O. Box 653, Beer-Sheva 84105, Israel, (phone: +972-8-6472458, fax: +972-8-6472903, e-mail: sokolovv@bgumail.bgu.ac.il).

I. Vajda, is with the Department of Electric Power Engineeing, Budapest University of Technology and Economics, Egry Jozsef utca 18, H-111 Budapest, Hungary. (phone: +36-1-4632904, fax: +36-1-4633600, e-mail: vajda@supertech.vgt.bme.hu).

G. Levin is with the Propulsion Directorate, Air Force Research Laboratory, Wright-Patterson Air Force Base, OH 45433 USA (e-mail: George.Levin@WPAFB.AF.MIL ).

We restrict ourselves to the consideration of an infinite long slab and a thin strip within the limits of the Bean model, and a coated conductor in the approximation of a strong external field.

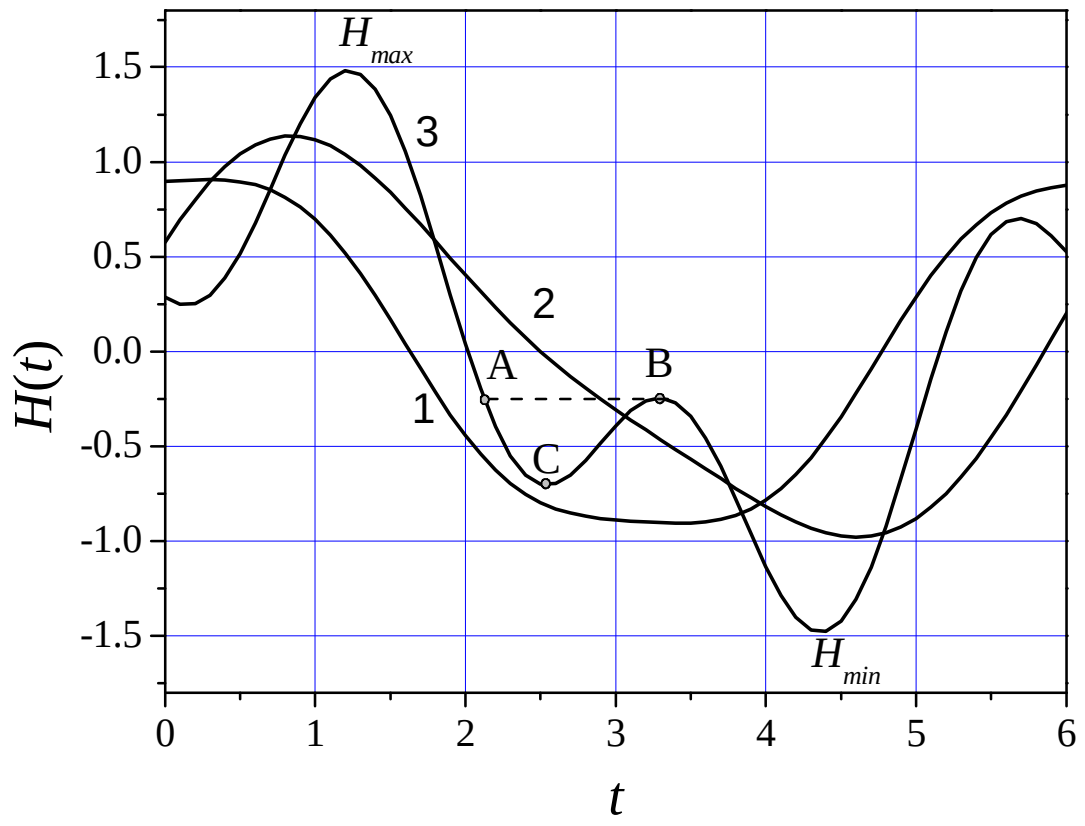

Fig. 1. Non-sinusoidal current waveform: 1 – symmetrical case; 2 - asymmetrical case; 3 - non-monotonic case.

## II. Loss estimation in the framework of Bean's model

The waveforms of non-sinusoidal currents (magnetic fields) can be divided into three types: I. symmetrical case - a current monotonically increases (decreases) from the minimum (maximum) to the maximum (minimum), and the maximum equals the module of the minimum (Fig. 1, line 1); II. asymmetrical case - a current monotonically changes like in the first case but the maximum does not equal the module of the minimum (Fig. 1, line 2); III. a current change is non-monotonic (Fig. 1, line 3). The type of a waveform depends on the number of harmonics, their frequencies and phases. If there are two harmonics $H(t) = H_1 \sin(\omega t) + H_k \sin(k\omega t)$, the monotonic behavior is observed at $H_k/H_1 < 1/k^2$.

It is known that if a current monotonically changes, in the framework of the Bean model the AC losses are determined only by the extreme values and do not depend on a current form. Using approaches developed in [1-4] we obtained the following expressions for the AC losses for the asymmetrical case II:

(a) losses per surface unit of a slab in a parallel magnetic field are

$$P = \begin{cases} \mu_0 f \dfrac{(H_{\max} - H_{\min})^3}{6 j_c}, & \text{if } |H_{\max} - H_{\min}| < 2H_p \\ 2\mu_0 f \dfrac{H_p{}^2 (H_{\max} - H_{\min} - \frac{4}{3} H_p)}{j_c}, & \text{if } |H_{\max} - H_{\min}| \geq 2H_p \end{cases} \quad (4)$$

(b) losses per length unit of a thin strip in a perpendicular magnetic field are

$$P = 4\mu_0 w^2 f J_c H_f \left( \frac{H_{\max} - H_{\min}}{2H_f} \right) g \left( \frac{H_{\max} - H_{\min}}{2H_f} \right) \quad (5)$$

(c) losses per length unit of a thin strip with a transport current are

$$P = \frac{\mu_0 f I_c^2}{\pi} h \left( \frac{I_{\max} - I_{\min}}{2I_c} \right), \quad (6)$$

where $h(x) = (1-x)\ln(1-x) + (1+x)\ln(1+x) - x^2$, $g(t) = \frac{2}{t} \ln \cosh t - \tanh t$, $H_{max}$, $H_{min}$, $I_{max}$, $I_{min}$ are the maximum and minimum of the magnetic field and current, respectively, $j_c$ is the critical current density, $J_c = j_c a$, $H_f = J_c/\pi$, $I_c = 2wJ_c$ is the critical current of the strip, $a$ and $w$ are the thickness and half-width of the strip, respectively, $H_p = j_c w_s$ is the complete penetration field, $w_s$ is a half of the slab thickness.

Note, that AC losses caused by a current in a slab can be estimated using the first equation of (4) where values of the magnetic field are replaced by the current per height unit of the slab.

As seen from Eqs. (4)-(6), the hysteresis losses are independent of frequency of the higher harmonics. These losses depend only on the change of the maximum and minimum of the current and/or magnetic field. Therefore, the AC loss calculation accounting the main harmonic only leads to an error determined by the difference $\Delta H = (H_{max} - H_{min}) - 2H_1$. To a first approximation, the AC losses can be estimated as

$$P \approx P_1 (1 + K \, \Delta H / 2H_1) \quad (7)$$

where $P_1$ is the AC losses caused by the main harmonic, $K$ is the coefficient depending on the form of a superconductor and on the value of $H_1$.

So, $K$ = 3 for a slab at $H_{max} - H_{min} < H_p$; $K$ = 4 for a strip at $H_{max} - H_{min} << H_f$ or at $I_{max} - I_{min} << I_c$; $K$ = 1 for both configurations at high magnetic fields. A relatively small difference $\Delta H/2H_1$ leads to a noticeable increase of the AC losses. For example, the losses increase by up to 20% at $\Delta H / 2H_1$ = 0.05. At the same time, in the normal metal, five-percent second harmonic causes the loss increase by 1%. The difference $\Delta H$ is determined not only by the harmonic amplitudes but also their phases (Fig. 2). And what is more, this difference can be negative (Fig. 2b) and, following, the appearance of higher harmonics can reduce AC losses. Designing the superconducting power devices, one should calculate AC losses for the worst case when the losses are maximal. For $k$ = 2, the maximum of $\Delta H$ is achieved at $\varphi_2 = 0$ and increases as about $2.4 H_2^{1.8}$, for $k$ = 3 this maximum equals $2H_3$ at $\varphi_3 = \pi$.

Consider a non-monotonic current change, when additional minima and maxima appear in the waveforms of magnetic field or current (Fig. 1, line 3). The field cyclically changes from the point A to the point B through C. The AC losses induced during this cycle are calculated using Eqs. (4)-(6) where $H_{max}$ and $H_{min}$ are replaced by $H_A$ and $H_C$, respectively. The number of "wells" and their depths, $H_A - H_C$, depend on the number of harmonics, their frequencies, amplitudes and phases. If $H(t) = H_1 \sin(\omega t) + H_k \sin(k\omega t + \varphi_k)$, the maximum well number is $k$-1. Thus, the total AC losses per the period are the sum of the losses contributed by all the cycles. Fig. 3 presents the characteristic values of a two-harmonic waveform for $\varphi_2 = \pi/2$. At $H_2/H_1 < 0.25$, $H_{max} - H_{min} = 2H_1$ and the losses are determined only by the main harmonic amplitude. If $H_2$ is higher, the losses increase due to the growth of $H_{max} - H_{min}$ and the contribution of an additional, "inner", cycle. If $H_A - H_C < 2H_p$, the last contribution achieves 5% at $H_2/H_1 = 1$; for $H_A - H_C >> 2H_p$ the 5% level is observed at $H_2/H_1 = 0.4$. So, one has to account the loss part appearing due to wells at relatively high amplitudes of higher harmonics.

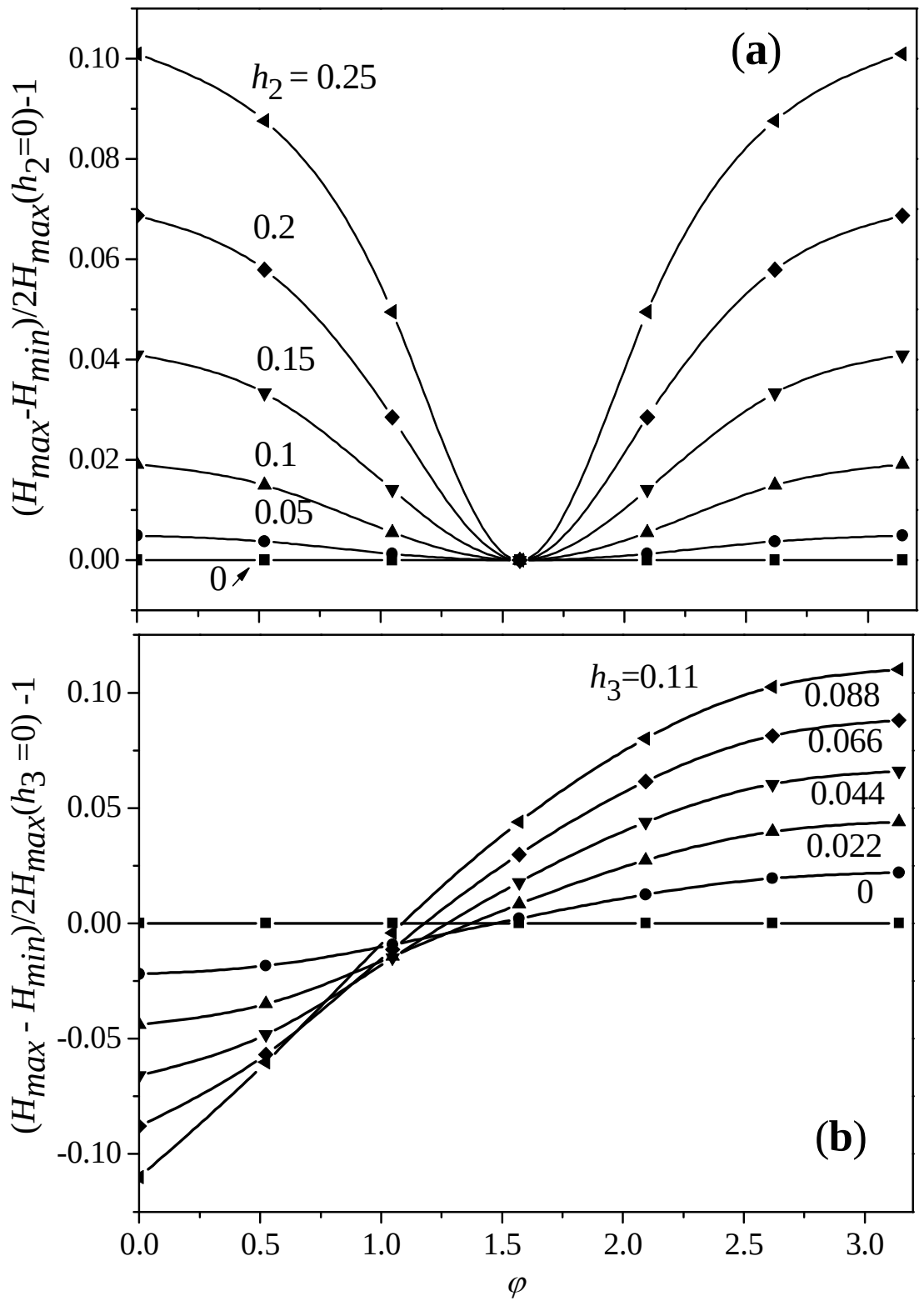

Fig. 2. Dependences of $\Delta H$ on phase for the second (a) and third (b) harmonics.

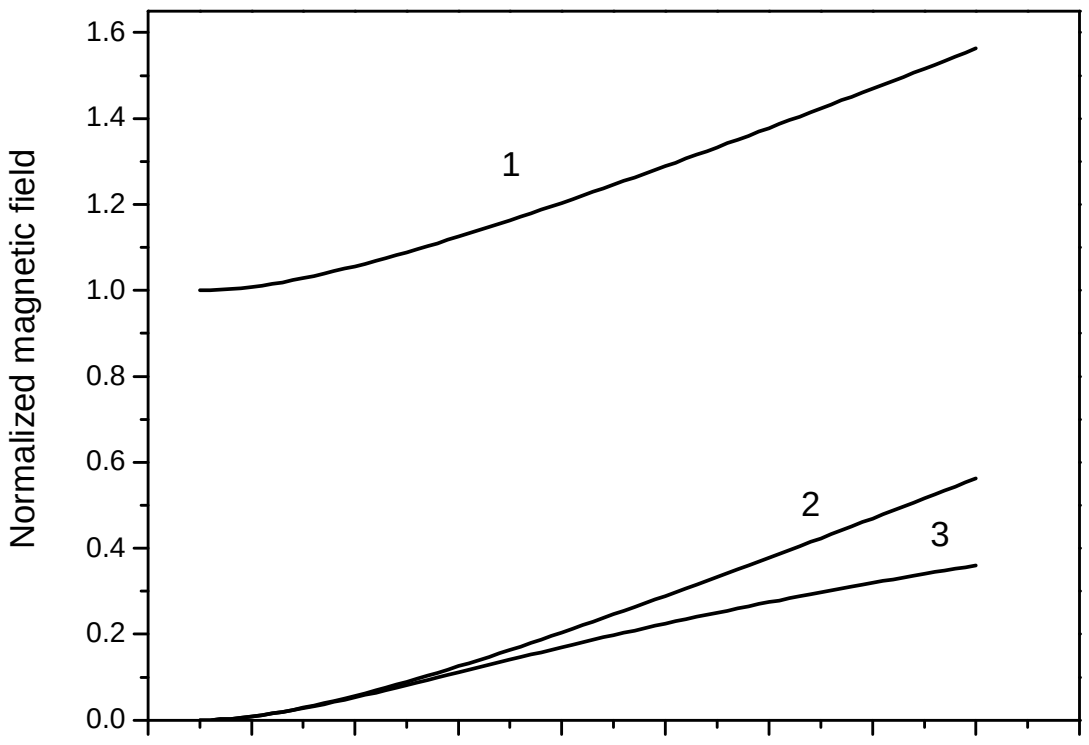

Fig. 3. Normalized characteristic values for $H(t)=H_1\sin(\omega t)+H_2\sin(2k\omega t+\pi/2)$ as functions of $H_2/H_1$: 1- $(H_{max} - H_{min})/2H_1$; 2 - $(H_A - H_C)/2H_1$; 3 - $(H_A - H_C)/(H_{max} - H_{min})$.

## III. Losses in coated conductors

Let us consider losses caused by a perpendicular magnetic field in a striped coated superconductor. It was shown [5] that the losses can be presented as a sum of the loss in a normal conducting substrate, the transverse coupling current loss and the hysteresis loss in the filamentary strips. Here we do not discuss the losses in the stabilizer. For a sufficiently high magnetic field these losses can be estimated separately assuming that the total magnetic field equals to the applied one. This approximation can be used for the loss estimation in magnetic fields with amplitudes higher than 0.06-0.6 T, typical for transformers, electrical motors and generators [6]. Under this assumption, we can calculate the coupling loss and loss in a substrate per length unit using Eq. (2b). Generalizing Eq. (44) from [5], we obtained for the losses $p_k = k^2$. The value of $K_1$ is given by Eq. (3) for the losses in the substrate, while for the coupling losses this coefficient is $w^2 L\omega^2 / 3(N-1) R_g$, where $L$ is the conductor length or the twisting step length, $N$ is the number of filamentary strips, $R_g$ is the resistance between two neighboring filaments.

Losses in superconducting strips can be estimated using the results obtained above using the Bean model. In this model, the linear current density cannot be higher than the critical value independent of a magnetic field. A more realistic picture is given by models based on a power law fitting of the voltage-current characteristic of a superconductor or/and taking into account the dependence of the critical current on a magnetic field. It was shown [6] that at high magnetic fields the loss values obtained for power-law dependence even with a high index of power (10-30) are markedly higher than those given by the critical state models.

Let us analyze the influence of higher harmonics on losses in a thin superconducting strip with the power-law relation between the current density $j$ and electric field $E$:

$$E = E_0 \left( \frac{j}{j_c} \right)^n, \qquad j_c = j_{c0} \frac{H_b}{|H| + H_b}, \tag{7}$$

where $E_0$ = 1 μV/cm is the electric field caused by the critical current density $j_c$ and $j_{c0}$ is its value at $H$ = 0, $H_b$ is a constant determined by the properties of the superconductor.

Losses per unit of the strip length are

$$P = P_0 \left[ \frac{n}{4n+2} \left( \frac{\mu_0 w \omega H_1}{E_0} \right)^{1/n} \right] F, \tag{8}$$

$$F = \int_0^T \frac{\omega \left| \sum_{k=1} h_k k \cos(\omega k t + \phi_k) \right|^{1+1/n}}{1 + h_b \left| \sum_{k=1} h_k \sin(\omega k t + \phi_k) \right|} dt$$

where $h_b = H_1/H_b$, $P_0 = 4 f \mu_0 j_{c0} w^2 H_1^2$ is the loss given by the Bean model in asymptotically high magnetic fields. The dependence of losses on the properties of a superconductor, $n$,

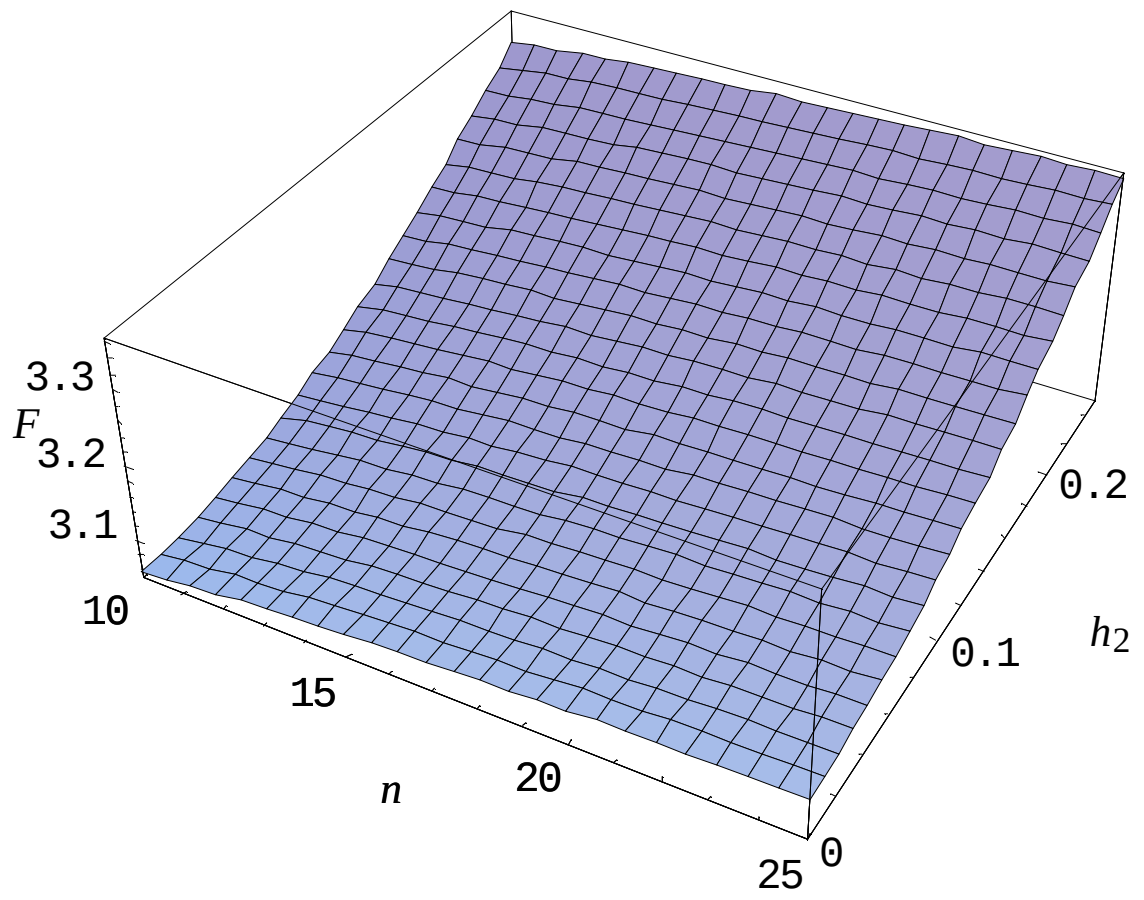

Fig. 4. Value of $F$ in (8) for $H(t) = H_1\sin(\omega t) + H_2\sin(2\omega t)$.

$E_0$, $H_b$, has been investigated in [6] for a sinusoidal magnetic field.

The influence of higher harmonics on losses appears in the value $F$ of the integral in (8). For a sinusoidal field ($h_k = 0$ for $k > 1$) and at $n\rightarrow\infty$, $h_b = 0$, (the Bean model), this integral equals 4 and $P = P_0$. Let us analyze the value $F$ for two cases of a two-harmonic waveform: (a) $k$ = 1, 2; (b) $k$ = 1, 3. $F$ has the maximum at $\varphi_2 = 0$ in the case (a) and $\varphi_3 = \pi$ in the case (b). The results of numerical integration for the first case are presented in Fig. 4 at $h_b = 0$ and $\varphi_2 = 0$. The value of $F$ decreases with an increase of $h_b$ approximately as $1/(1+0.85h_b)^{0.625}$ for both cases and increases with the amplitudes of higher harmonics proportional to $1+a_2h_2^{1.8}$ for the first case and $1+a_3h_3$ for the second case. The coefficients $a_2$ and $a_3$ depend on the power index $n$ and parameter $h_b$. For $h_b < 1$ and $n > 10$, within the accuracy of 10%, $a_2$ and $a_3$ can be taken 1.3 and 0.9, respectively. The increase of $n$ above 20 does not lead to a marked growth (the difference of the values of $F$ at $n$ = 20 and $n$ = 40 is less than 2%).

## IV. Conclusion

Our results show that in the systems with superconducting elements, higher harmonics can substantially change the amount of power loss. While, in the normal metal, the 5% second harmonic causes the loss increase by 1%, in superconductors this increase can achieve 20%. Moreover, the contribution of the harmonics depends on their phases: in a certain range of phases, the odd harmonics can even reduce AC losses. These peculiarities distinguish the behavior of superconducting devices from that of conventional ones.

Non-monotonic character of the waveforms of a current or a magnetic field influences AC losses only at very high amplitudes of additional harmonics.